\documentclass[preprint,12pt]{revtex4}

\usepackage[brazil, english]{babel}
\usepackage[utf8]{inputenc}
\usepackage{graphicx}
\usepackage{epsfig}
\usepackage{amssymb}
\usepackage{amsmath} 
\usepackage{slashed}
\usepackage[unicode=true,bookmarks=false,breaklinks=false,pdfborder={0 0 1},colorlinks=true]
 {hyperref}
\hypersetup{
 citecolor=blue,linkcolor=blue,urlcolor=blue}

\graphicspath{%
    {converted_graphics/}
    {/}
}
\begin{document}

\title{Fluctuation and dissipation within a\\ deformed holographic model with backreaction} 

\author{Nathan G. Caldeira$^{1,}$}
\email[Eletronic address: ]{nathangomesc@hotmail.com}
\author{Eduardo Folco Capossoli$^{1,2,}$}
\email[Eletronic address: ]{eduardo\_capossoli@cp2.g12.br}
\author{Carlos A. D. Zarro$^{1,}$}
\email[Eletronic address: ]{carlos.zarro@if.ufrj.br}
\author{Henrique Boschi-Filho$^{1,}$}
\email[Eletronic address: ]{boschi@if.ufrj.br}  
\affiliation{$^1$Instituto de F\'{\i}sica, Universidade Federal do Rio de Janeiro, 21.941-972 - Rio de Janeiro - RJ - Brazil \\ 
 $^2$Departamento de F\'{\i}sica and Mestrado Profissional em Práticas de Educa\c{c}\~{a}o B\'{a}sica (MPPEB), 
 Col\'egio Pedro II, 20.921-903 - Rio de Janeiro - RJ - Brazil\\
}

\begin{abstract}
{ In this work we study the fluctuation and dissipation of a string attached to a brane in a deformed and  backreated AdS-Schwarzschild spacetime. This space is a solution of Einstein-dilaton equations  and contains a conformal exponential  factor $\exp(k/r^2)$  in the metric. We consider the backreaction contributions coming only from the exponential warp factor on the AdS-Schwarzschild black hole, where the string and brane are in the probe approximation.  
  Within this Lorentz invariant holographic model we have computed the admittance,  the diffusion coefficient, the two-point functions and the regularized mean square displacement $s^2_{reg}$. From this quantity we obtain the diffuse and ballistic regimes characteristic of the Brownian motion. From the two-point functions and the admittance, we also have checked the well know fluctuation-dissipation theorem in this set up.}

\end{abstract}


\maketitle


\section{Introduction}

Brownian motion \cite{brown, lange}  and the  fluctuation-dissipation theorem \cite{kubo} stand until today as two of the most important subjects within non-equilibrium statistical mechanics. Its intersections and contributions spread over many branches of science and in particular at high energy physics, such as matter under extreme conditions or the quark-gluon-plasma (QGP) \cite{Policastro:2001yc, CasalderreySolana:2011us}. In this case, the constituents of nuclear matter, under high temperature or density, present erratic trajectories due to their interactions with each other behaving like a Brownian motion. In this sense, by studying QGP one can investigate those phenomena. A very interesting approach to deal with non-perturbative aspects of strong interactions, which appear in such processes, is based on the AdS/CFT correspondence \cite{Aharony:1999ti} which relates a weak coupled theory in a curved spacetime (AdS$_5$) with a strong coupled theory in four dimensional Minkowski spacetime. An incomplete list of references which dealt with Brownian motion, dissipation, fluctuation and related topics  resorting to the AdS/CFT correspondence can be found, for instance, in Refs. \cite{deBoer:2008gu, Son:2009vu, Atmaja:2010uu, Tong:2012nf, Edalati:2012tc, Fischler:2014tka, Giataganas:2013hwa, Giataganas:2018ekx}.

In Ref. \cite{Caldeira:2020sot} the authors studied the fluctuation and the dissipation through an AdS/QCD model based on a deformation of the AdS-Schwarzschild spacetime. This deformation is due to the introduction of a  conformal factor $\exp({k}/{r^{2}})$ in the metric of such a space. Then they computed the string energy, the response function, the mean square displacement, the diffusion coefficient and checked the fluctuation-dissipation theorem. This and related deformed AdS/QCD models were used successfully in many holographic problems as can be seen, for example, in Refs. \cite{Andreev:2006vy, Rinaldi:2017wdn, Bruni:2018dqm, Afonin:2018era, FolcoCapossoli:2019imm, FolcoCapossoli:2020pks}.

Here, in this work we will use a probe string attached to a probe brane in a Lorentz invariant deformed AdS/QCD model taking into account the backreaction from the exponential factor on the metric of the AdS-Schwarzschild space. This will allow us to extend the work done in Ref. \cite{Caldeira:2020sot} and investigate the contribution of the backreaction on the admittance, the diffusion coefficient, the mean square displacement and the fluctuation-dissipation theorem in this set up. 
In previous studies \cite{Son:2009vu, Atmaja:2010uu, Tong:2012nf, Edalati:2012tc, Fischler:2014tka, Giataganas:2013hwa, Giataganas:2018ekx}, the backreaction was not considered. 

This work is organized as follows: In Sec. \ref{emd} we present the Einstein-dilaton action and solve the corresponding field equations. From these field equations we obtain the backreacted horizon function consitent with the deformed warp factor and the dilaton potential. In Sec. \ref{bulkdesc} we describe our holographic model and study the effects of the backreaction on the fluctuation and dissipation at the string endpoint attached to the brane. In Sec. \ref{meanflu} we compute the mean square displacement from which we obtain the ballistic and the diffusive regimes characteristic of the Brownian motion. We also check the fluctuation-dissipation theorem in this set up. Finally, in Sec. \ref{conc} we present final discussions and our conclusions.

\section{The Einstein-Dilaton action and the deformed AdS-Schwarschild space with backreaction} \label{emd}

In order to capture all features of our deformed and backreacted space, let us start with a 5-dimensional Einstein-dilaton action in Einstein frame:
\begin{equation} \label{EMD action}
S = \dfrac{1}{16\pi G_5}\int d^{5}x \sqrt{-g}\left(R - \dfrac{4}{3}g^{\mu\nu}\partial_{\mu}\phi\partial_{\nu}\phi + \mathcal{V}(\phi) \right),
\end{equation}
where $ G_5 $ is the 5-dimensional Newton's constant, $g$ is the metric determinant, $ R $ is the Ricci scalar, $\phi$ is the dilaton field and $\mathcal{V}(\phi)$ its potential. From this action one obtains the field equations
\begin{eqnarray}
G_{\mu\nu} -\dfrac{4}{3}\left(\partial_{\mu}\phi\partial_{\nu}\phi - \dfrac{1}{2}g_{\mu\nu}(\partial \phi)^2\right) - \dfrac{1}{2}g_{\mu\nu}\mathcal{V}(\phi) &=& 0, \label{EinsteinEqn} \label{eom1}\qquad \\ 
\nabla^{2}\phi + \dfrac{3}{8}\frac{\partial\,\mathcal{V}(\phi )}{\partial\,\phi}&=& 0, \label{DilatonEqn} \label{eom2}
\end{eqnarray}
where $G_{\mu\nu} = R_{\mu\nu}- \dfrac{1}{2}g_{\mu\nu}R$ is the Einstein tensor. For our purposes, as done in Refs. \cite{Ballon-Bayona:2017sxa, Ballon-Bayona:2020xls}, we will consider the ansatz
\begin{eqnarray}
ds^2 &=& \dfrac{1}{\zeta(z)^2}\left(\dfrac{dz^2}{f(z)} - f(z)dt^2 + d\vec{x}^2\right), \label{Metriczeta} \end{eqnarray}
\noindent where $z$ is the holographic coordinate, $f(z)$ is the horizon function and $\zeta(z)$ is the metric warp factor which we choose to be
\begin{equation}\label{warp}
  \zeta(z) = z \; e^{-\frac{1}{2} \left(k z^2\right)} \,,
\end{equation}
\noindent with $k$ being the deformation parameter which plays the role of the IR scale in the model. 

Replacing the ansatz, Eq. \eqref{Metriczeta},  into the Einstein-dilaton equations  of motion \eqref{eom1} and \eqref{eom2} one gets: 
\begin{eqnarray}
\frac{d}{dz}\left(\zeta(z)^{-3} \frac{d}{dz} f(z)\right) &=& 0\,,  \label{freqn3} \\ 
\frac{\zeta''(z)}{\zeta(z)} - \frac{4}{9}\phi'(z)^2 &=& 0,\label{breqn3} 
\end{eqnarray}
where $' \equiv d/dz$. By using Eq. \eqref{eom2} one can obtain an expression  for the dilaton potential, so that:
\begin{eqnarray}  \label{potential}
{\cal V}(\phi) = 12\zeta'(z)^2 f(z) - 3\zeta'(z)f'(z)\zeta(z) - \frac{4}{3}f(z)\zeta (z)^2\phi'(z)^2 \;\;, 
\end{eqnarray}
and from Eq. \eqref{breqn3} we get:
\begin{equation}\label{dil}
    \phi(z)=c_1\pm\frac{3}{4} \sqrt{k \left(k z^2-3\right)} \left(z-\frac{3 \log \left(\sqrt{k(k z^2-3)}+k z\right)}{\sqrt{k(k z^2-3)}}\right)\,, 
\end{equation}
where $c_1$ is an integration constant.  Substituting Eq. \eqref{warp} into Eq. \eqref{freqn3}, satisfying  $f(0) = 1$ and the horizon property $f(z_h) = 0$, one can solve it analytically, so that: 
\begin{align}\label{horfunfinal}
f(z)=1&-\left(\frac{3 k z^2-2 e^{\frac{3}{2} k z^2}+2}{3 k z_{h} ^2-2 e^{\frac{3}{2} k z_{h} ^2}+2}\right) \; e^{\frac{3}{2} k (z_{h}^2- z^2)}\,. \end{align}
This is the horizon function with backreaction coming from the exponential deformation in the metric, disregarding contributions from the probe string or the probe brane. These probes will be introduced in the following.

One can verify that the AdS-Schwarzschild space is recovered in the limit $k \to 0$ for which $f^{\rm AdS-Sch}(z) =  1 -z^4/z_h^4$  and the potential ${\cal V}(\phi)$ reduces to a constant ${\cal V}(\phi) = 12$ with AdS radius $L=1$. The 
 Eq. \eqref{horfunfinal} also fulfills the condition $f'(z_h)<0$. 
 
If one substitutes the warp factor, Eq. \eqref{warp}, the dilaton profile in Eq. \eqref{dil} and the horizon function, Eq. \eqref{horfunfinal} into the expression of the potential \eqref{potential}, one can obtain an  expression for the potential in terms of the $z$ coordinate. For simplicity, we are not presenting this expression explicitly.

\begin{figure}[ht]
	\centering
	\includegraphics[scale = 0.4]{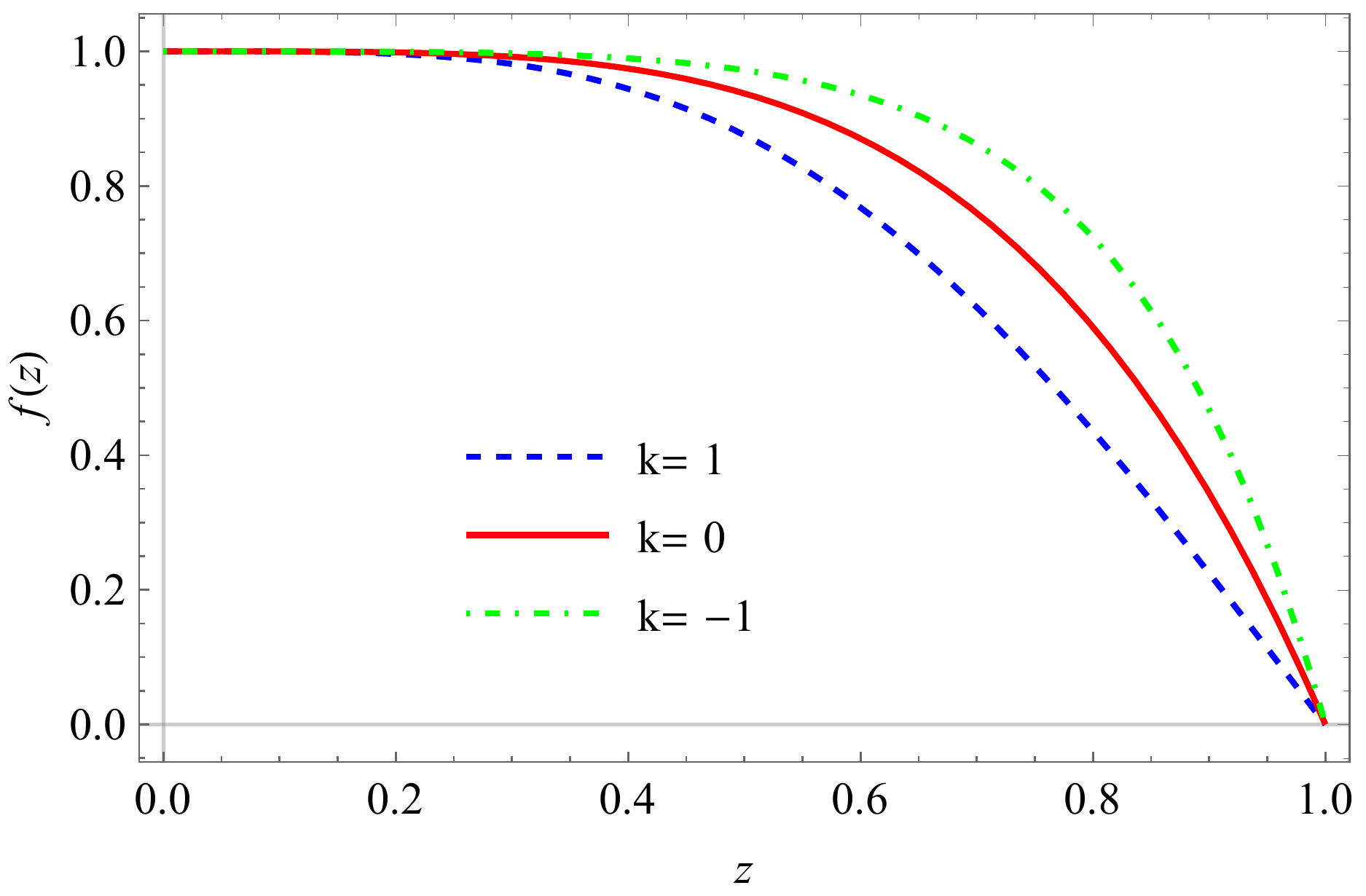}
	\caption{The horizon function $ f(z) $, Eq. \eqref{horfunfinal}, vs the holographic coordinate $z$  for $k=0$ and $k \pm1$ and  $z_h = 1$ in arbitrary  units.}
	\label{hor_mu_zero}
\end{figure}

 In Figure \ref{hor_mu_zero}, we present the behavior of the horizon function in terms of the holographic coordinate $z$ for both signs of the constant $k$.

\section{Probe string in the bulk with backreaction effects}\label{bulkdesc}
In this section we implement the description of a probe string attached to a probe brane in a thermal bath with backreaction from the exponential deformation of the metric. For convenience, we change the coordinate $z$ to 
 $r=1/z$ so that the metric, Eq. \eqref{Metriczeta}, is rewritten as 
\begin{equation}\label{metrictemp}
    ds^2 = e^{\frac{k}{r^2}} \left[-r^{2}f(r)dt^{2}+{r^2}\left(\eta_{i j} dx^{i}dx^{j} \right) +\frac{dr^2}{r^{2}f(r)}\right].
\end{equation}
Also in $r$ coordinate, the horizon function, Eq. \eqref{horfunfinal}, reads:
\begin{equation}\label{hormuzeror}
    f(r) = 1-\left(\frac{\frac{3 k}{r^2}-2 e^{\frac{3 k}{2 r^2}}+2 }{\frac{3 k}{r_{h}^2}-2 e^{\frac{3 k}{2 r_{h}^2}}+2}
    \right)
    e^{-\frac{3}{2} k \left(\frac{1}{r^2}-\frac{1}{r_{h}^2}\right)}\,. 
\end{equation}

In order to describe the string we consider the Nambu-Goto action, given by $S_{NG} = - \frac{1}{2 \pi \alpha'} \int d\tau d\sigma \sqrt{-\gamma}$, where $\alpha'$ is the string tension, $\gamma = {\rm det} (\gamma_{\alpha \beta})$ and $\gamma_{\alpha \beta} = g_{mn} \partial_{\alpha}X^m \partial_{\beta}X^n $ is the induced metric on the worldsheet with $m,n = 0, 1, 2, 3, 5$. We also choose a static gauge, where $t = \tau$, $r = \sigma$ and $X= X(\tau, \sigma)$. Expanding the Nambu-Goto action, keeping only the quadratic terms $\dot{X}^2$, $X'^2$, and using the metric, Eq. \eqref{metrictemp}, we get:  
\begin{equation}\label{ngapprox}
S_{NG} \approx - \frac{1}{4 \pi \alpha'} \int dt dr \left[ \;\dot{X}^2 \frac{e^{\frac{ k}{r^2}}}{f(r)}-X'^2 r^4 f(r) e^{\frac{k}{r^2}} \right]\,,
\end{equation}
\noindent where $\dot{X}=\partial_{t} X$ and $X'=\partial_{r} X$. Factorizing $X(t,r)$ as  $X(t,r)=e^{i\omega t}h_{\omega}(r)$, the equation of motion reads:
\begin{equation}\label{EquationofMotionSecondVersion}
    \frac{\partial }{\partial r}\left(r^{4}f(r)e^{\frac{k}{r^{2}}}h_{\omega}'(r,t)\right)-\frac{e^{\frac{k}{r^{2}}}\omega^{2}}{f(r)}h_{\omega}(r)=0.
\end{equation}

Going to the tortoise coordinate 
    $ r_{*}=\int {dr}/\left({r^{2}f(r)}\right)$ 
\noindent and making a Bogoliubov transformation $ h_{\omega}(r_{*})=e^{B(r_*)}\psi(r_*)$, where $B(r)= -k/{2 r^2}-\log (r)$, we obtain a Schrödinger-like equation:
\begin{equation}\label{sch}
    \frac{d^{2}\psi(r_{*})}{dr_{*}^{2}}+\left(\omega^{2}-V(r)\right)\psi(r_{*})=0,
\end{equation}
with potential 
\begin{eqnarray}
    V(r)&=&-f(r)\left(\left(-\frac{k^2}{r^2}+k-2 r^2\right) f(r)+r \left(k-r^2\right) f'(r)\right)\,. 
\end{eqnarray}
As the equation \eqref{sch} cannot be analytically solved, we will apply the monodromy patching procedure \cite{deBoer:2008gu, Caldeira:2020sot} and seek for approximate analytical solutions. For our purposes we will choose three regions: \textbf{A}, \textbf{B}, \textbf{C} and explore their solutions. First, we consider the region {\bf A} which is near the horizon ($r\sim r_{h}$). In this region one has $V(r)\ll\omega^{2}$, so that the Schrödinger equation \eqref{sch} reads
\begin{equation}\label{schsemw}
    \frac{d^{2}\psi(r_{*})}{dr_{*}^{2}}+\omega^{2}\psi(r_{*})=0,
\end{equation}
which has the ingoing solution 
$ \psi(r_{*}) = A_1 e^{-i\omega r_*}.$ 
For low frequencies one can expand this  solution as $ \psi(r_{*}) = A_1  -i A_1 \omega r_* $ allowing us to compute $h_{\omega}(r_{*})$ in this region:
\begin{equation}\label{hAr*}
h_{\omega}^{A}(r_*)=A_1\frac{e^{-\frac{k}{2 r^2_h}}}{r_h}\left(1- i\omega\lambda  \log \left(\frac{r}{r_{h}}-1\right) \right)\,,
\end{equation}
\noindent where we defined the quantities
\begin{equation}
     \lambda \equiv  \frac{2\left(e^{\frac{3}{2}x}-1\right)-3x}{9x^{2}r_{h}} \,\, ; \qquad x\equiv k/r_{h}^{2}\,.
\end{equation}

The next region, {\bf B}, is defined as  $V(r)\gg\omega^{2}$. In this region the equation of motion  \eqref{EquationofMotionSecondVersion} becomes:
\begin{equation}
\label{reg2}
    \frac{d }{d r}\left(r^{4}f(r)e^{\frac{k}{r^{2}}}h_{\omega}'\right)=0\,.
\end{equation}

The solution of the IR part of this region can be written as
\begin{eqnarray}
    h^{B}_{\omega (\rm IR)}(r) \approx B_{1} \frac{\lambda}{r_{h}^{2}}e^{-x}\log\left(\frac{r}{r_{h}}-1\right)+B_{2}. 
\end{eqnarray}
where $B_1$ and $B_2$ are constants. Comparing with \eqref{hAr*} we get
\begin{eqnarray}
    B_{1}=-i A_{1} r_{h} \, \omega \, e^{\frac{k}{2 r_{h} ^2}}, \qquad 
    B_{2}=\frac{A_{1}}{r_{h}}e^{-\frac{k}{2r_{h}^{2}}}.
\end{eqnarray}

On the other hand, the solution of the UV part of this region  can be approximated by
\begin{equation}
\label{HBUV}
    h^{B}_{\omega(\rm UV)}(r)\approx -\frac{B_{1}}{3 r^3}+B_{2}.
\end{equation}

The last region, {\bf C}, represents the deep UV meaning that the horizon function reduces to $f(r) =1$. In this case, Eq. \eqref{EquationofMotionSecondVersion} has the  solution:
\begin{equation}\label{hwc}
h^C_{\omega}(r)=C_1\,  {}_1F_1\left(\frac{\omega^{2}}{4k},-\frac{1}{2},-\frac{k}{r^{2}}\right)+ C_2 \frac{(-k)^{3/2}}{r^3}{}_1F_1\left(\frac{3}{2} + \frac{\omega^{2}}{4k},-\frac{5}{2},-\frac{k}{r^{2}}\right)\,, 
\end{equation}
where ${}_1F_1(a,b,z)$ is the confluent hypergeometric function of the first kind. 
Close to the boundary, keeping only terms up to $O(\omega)$, it can be expanded as 
\begin{eqnarray}
   h^{C}_{\omega}(r)\approx C_1+\frac{C_2 k^{3/2}}{r^3}.
\end{eqnarray}

Matching $h^{B}_{\omega(\rm UV)}(r)$ and $h^{C}_{\omega}(r)$, one can write the solution close to the boundary (the brane) as
\begin{eqnarray}
    h^{C}_{\omega}(r)\approx \frac{A_1}{{r_{h}}}
    \left(e^{-\frac{k}{r_{h}^{2}}}
    +i \omega\frac{ r_{h}^2 } {3r^3}\right) e^{\frac{k}{2 r_{h} ^2}}\,. 
\end{eqnarray}
From this solution one can calculate the linear response or the admittance $\chi(\omega)$ of the string endpoint on the brane. Such a response is due to the action of an external force in an arbitrary brane direction, $x^{i}$, and can be written as $F(t) = E \, e^{-i\omega t} F(\omega)$, where $E$ is the electric field on the brane. Following Refs. \cite{ Tong:2012nf, Caldeira:2020sot} one can write the force as:
\begin{eqnarray}
 F(\omega)=\frac{A_{1}}{2 \pi \alpha'}\left[-i \omega r_{h} e^{\frac{k}{2 r_{h} ^2}} f(r_{b})e^{\frac{ k}{r_b^2}}\right]\,. 
\end{eqnarray}

Considering the limits where the brane is far away from the horizon $r_{b}\gg r_{h}$, and the scale of the probe brane is much greater than the IR scale $r_{b}\gg \sqrt{k}$, then $f(r_{b})\to 1$, therefore the admittance, is given by:
\begin{eqnarray}
\label{Admittance}
 \chi(\omega)\equiv \frac{h^{C}_{\omega}(\omega)}{F(\omega)}
=\frac{2\pi i\alpha'}{\omega r_{h}^{2}} e^{-x}=\frac{2\pi i\alpha'}{\omega g_{ ii}(r_{h})}, \,\,\,{\rm with}\,\,\,x\equiv k/r_{h}^{2}\,. 
\end{eqnarray}
Note that the last equality in the above equation was proposed in Ref. \cite{Giataganas:2018ekx}, in the context of a general polynomial metric, where  $g_{ii}$ is the metric component in the $x^i$ direction. Following this reference, one  can write $\chi(\omega)$ as:
\begin{equation}
     \chi(\omega)=2\pi\alpha'\left(\frac{i}{\gamma \omega} - \frac{\Delta m}{\gamma^{2}} +\mathcal{O}(\omega)\right),
\end{equation}
\noindent with  
\begin{align}
    \gamma=e^{\frac{k}{r_{h}^{2}}}r_{h}^{2}\left(1+\frac{k}{r_{b}^2}+O\left(\frac{1}{r_{b}^{3
    }}\right)\right), && \Delta m=\frac{e^{-\frac{3k}{2r_{h}^{2}}}r_{h}^{4}}{r_{b}^{3}}\left(1+O\left(\frac{1}{r_{b}^{2}}\right)\right), 
\end{align}
where $\gamma$ is the friction coefficient and $\Delta m$ corresponds to the change in the bare mass $m$ of the particle in the Langevin equation \cite{lange,kubo}. 

The Hawking temperature associated with the black hole in our deformed AdS-Schwarzschild space, is given by:
\begin{eqnarray}
 \label{HawkingTemperature}
 T&=&\frac{r^{2}}{4\pi}\left|\frac{d f(r)}{d r}\right|\Bigg|_{r=r_{h}} = \frac{r_{h}}{\pi}g(x)\,,\,\,\quad  {\rm where} \,\,\, g(x) \equiv \left|\left(\frac{9 x^2}{4\left(2 \left( e^{\frac{3}{2}x}-1\right
 )-3x \right)}\right)\right|\,. 
\end{eqnarray}
It is worthwhile to mention that in the limit $k\to 0$ (or equivalently $x\to 0$), one recovers the AdS-Schwarzschild meaning  $T\to r_{h}/\pi$.

By using the above definition of the Hawking temperature in the expression for the admittance, Eq. \eqref{Admittance}, one can rewrite it as: 
\begin{equation}\label{chi}
    \chi(\omega)=\frac{2 i\alpha'}{\omega \pi T^{2}}\, e^{-x} \, g(x)^{2}.
\end{equation}

\begin{figure}
	\centering
	\includegraphics[scale = 0.38]{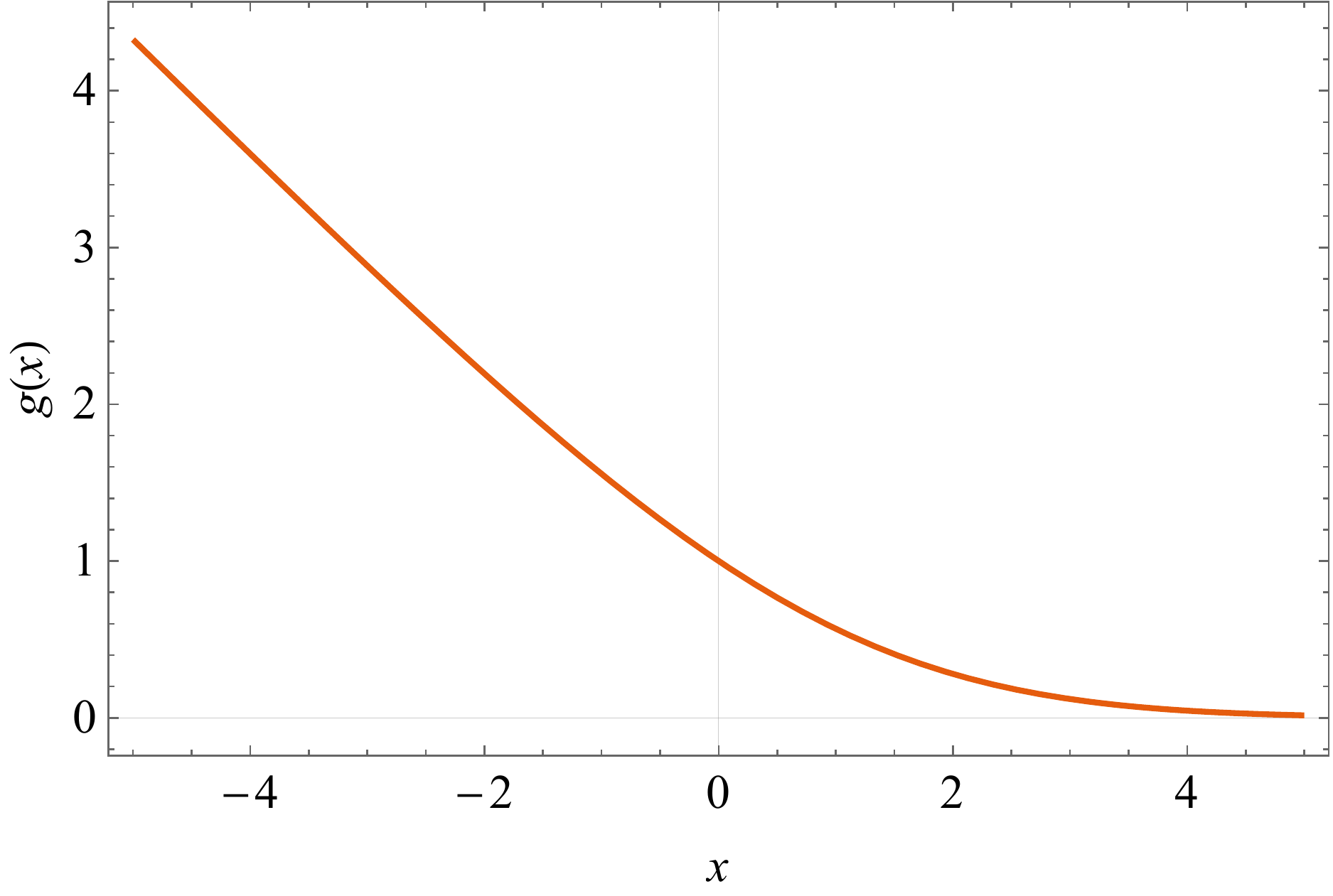}
	\hfill
	\includegraphics[scale = 0.38]{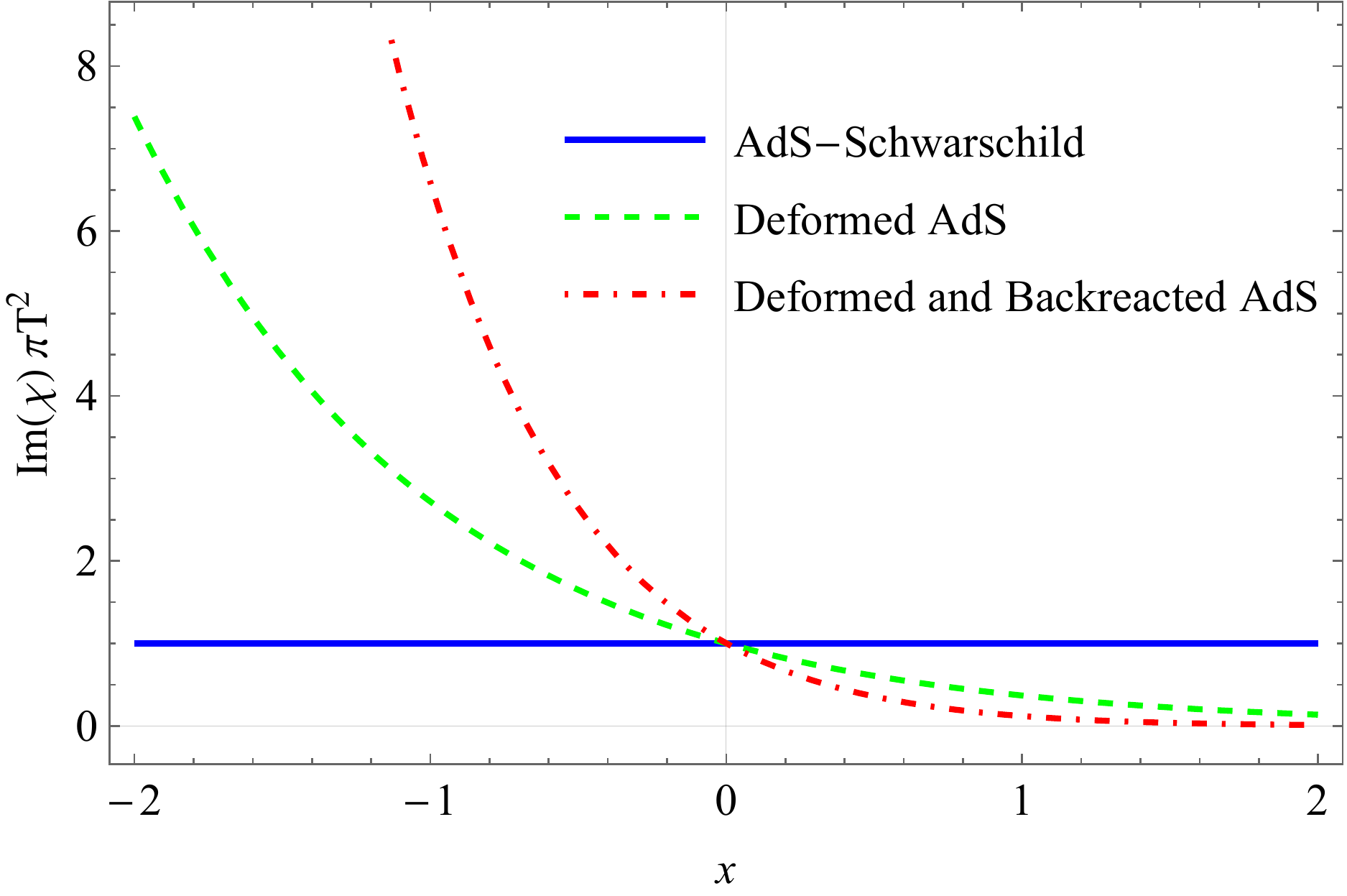}
	\caption{{\sl Left panel:} Plot of the  function $g(x)$ against $x=k/r_h^2$ that measures the shift from AdS-Schwarzschild Hawking temperature after backreaction.  {\sl Right panel:} Imaginary part of the admittance $\chi$ times $\pi T^2$ against $x$, for three situations: pure AdS-Schwarzschild, deformed AdS-Schwarzschild \cite{Caldeira:2020sot}, and deformed AdS-Schwarzschild with backreaction. Note the asymmetry between positive and negative values of $k$ (or $x$) in both panels.}
	\label{fig:resumo}
\end{figure}

At this point it is interesting to compare our result for the admittance with Ref. \cite{Caldeira:2020sot}, which computes this quantity for a deformed AdS-Schwarzschild space with no backreaction ($\chi_{_{NBR}}(\omega)$), and Refs. \cite{Tong:2012nf, Edalati:2012tc, Giataganas:2018ekx} where the authors compute the admittance in a geometry which includes the pure AdS-Schwarzschild ($\chi_{_{AdS}}(\omega)$). Note that: 
\begin{equation}\label{comparison}
    \chi(\omega)=\chi_{_{AdS}}(\omega)\, e^{-x} \, g(x)^{2}=\chi_{_{NBR}}(\omega)g(x)^{2}.
\end{equation}
In Fig. \ref{fig:resumo} we present the behavior of $g(x)$ and compare the admittances presented in Eq.  \eqref{comparison}.

Using the result for the admittance found here we can calculate the diffusion coefficient, which is given by
\begin{equation}
\label{DCoefficient}
  D = T \lim_{\omega \to 0}(- i \omega \chi(\omega))=\frac{2\alpha'}{\pi T}\, e^{-x}\, g(x)^{2} \,,
\end{equation}
where one can clearly see the contributions from the deformation ($e^{-x}$) of the AdS-Schwarzschild metric and the backreaction ($g(x)^{2}$), analogously to the admittances discussed above. It is worthwhile to mention that $x$ and $g(x)$ are implicit functions of the temperature $T$ which cannot be analytically inverted and then all the results presented are involved functions of the temperature, as can be seen, for instance, in Eqs. \eqref{chi} and \eqref{DCoefficient}.

\section{Mean Square Displacement and Fluctuation-Dissipation Theorem}\label{meanflu}

The Schr\"odinger equation \eqref{schsemw} has as solution a linear combination between the ingoing and outgoing modes. Considering the outgoing mode as $ \psi^{out}(r)=A_{2}e^{i\omega r_{*}}$ one can follow the above steps of the monodromy patching procedure and obtain  for the region {\bf A} an expression given by:
\begin{equation}
    h_{\omega}(r)=A\frac{e^{-\frac{k}{2r_{h}^{2}}}}{r_{h}}\left[\left(1
    +\frac{i \omega r^{2}_{h} e^{\frac{k}{ r_{h} ^2}} }{3r^3}\right)+B\left(1
    -\frac{i \omega r^{2}_{h} e^{\frac{k}{ r_{h} ^2}} }{3r^3}\right)\right].
\end{equation}

Similarly, for the region {\bf C} in terms of ingoing and outgoing modes, one has: 
\begin{eqnarray}
    h^{C}_{\omega}&=&A[h^{out}_{\omega}(r)+Bh^{in}_{\omega}(r)]\nonumber\\
        &=&A\left[ \,_1F_1\left(\frac{\omega ^2}{4 k };-\frac{1}{2};-\frac{k }{r^2}\right)-i\omega \frac{r^{2}_{h}  }{3r^3} \,_1F_1\left(\frac{\omega ^2}{4 k }+\frac{3}{2};\frac{5}{2};-\frac{k }{r^2}\right)
        e^{\frac{k}{r_{h} ^2}}
        \right.\nonumber\\
    & &\left.+ B\left( \,_1F_1\left(\frac{\omega ^2}{4 k };-\frac{1}{2};-\frac{k }{r^2}\right)+ i\omega \frac{ r^{2}_{h}   }{3r^3} \,_1F_1\left(\frac{\omega ^2}{4 k }+\frac{3}{2};\frac{5}{2};-\frac{k }{r^2}\right)
    e^{\frac{k}{r_{h} ^2}}\right)  \right]\,,
\end{eqnarray}
\noindent where $A$ and $B$ are constants to be determined. 

On the other hand, close to the horizon one can write the general solution as:
\begin{equation}
    h_{\omega}(r)=A\frac{e^{-\frac{k}{2r^{2}}}}{r}[e^{i\omega \lambda\log\left(\frac{r}{r_{h}}-1\right)}+Be^{-i\omega \lambda\log\left(\frac{r}{r_{h}}-1\right)}]\,,\quad {\rm  where}\;\; \lambda = \frac{1}{4 r_h g(x)}\,.
\end{equation}
\noindent Following \cite{deBoer:2008gu}, by imposing Neumann boundary conditions at the brane ($r=r_{b}$) and at the horizon where  $r/r_{h}=1+\epsilon$, with $\epsilon\ll 1$, one can write

\begin{equation}
    B=-\frac{h_{\omega}^{'out}(r)}{h_{\omega}^{'in}(r)}\Bigg|_{\frac{r}{r_{h}}=1+\epsilon} 
    \approx e^{-2 i\omega \lambda\log\left({1}/{\epsilon}\right)}\,,
\end{equation}
\noindent which produces discrete frequencies,  $\Delta\omega=\pi/\lambda\log\left({1}/{\epsilon}\right)$.

In order to compute the constant $A$ one can use the normalized Klein-Gordon inner product \cite{Giataganas:2018ekx, Caldeira:2020sot}
\begin{align}
    (X_{\omega}(r,t),X_{\omega}(r,t))
    &=\frac{\omega}{2\pi \alpha'}\int_{r_{h}}^{r_{b}}dr\;\frac{e^{\frac{k}{r^{2}}}}{f(r)} |h_{\omega}(r)|^{2}=1. \label{IKGP}
\end{align}
This integral is dominated by the near horizon region: 
%
\begin{eqnarray*}
\frac{2\omega|A|^{2}}{\pi \alpha'}\int_{r_{h}(1+\epsilon)}\frac{dr}{r^{2}f(r)}\approx \frac{2\omega\lambda |A|^{2}}{\pi \alpha'}\log\left(\frac{1}{\epsilon}\right)
\end{eqnarray*}
so that 
\begin{equation}
  A=\sqrt{\frac{\pi \alpha'}{2\omega\lambda \log\left({1}/{\epsilon}\right)}}\,.
\end{equation}

To compute the mean square displacement of the string endpoint located at the brane one has to write the thermal two-point function as a Fourier series:
\begin{equation}\label{fourrier}
X(t,r) = \sum_{\omega>0} \left(h^C_{\omega }(r)e^{-\text{i$\omega t $}} a_{\omega} + h^{C*}_{\omega}(r)e^{\text{i$\omega t $}}   a^{\dagger}_{\omega}\right) \, ,
\end{equation}
where the frequencies $\omega$ are discrete while $a_{\omega}$ and $a^{\dagger}_{\omega}$ are the annihilation and creation operators, respectively. Then,  disregarding terms of the order $1/r_b$ or less, one gets 
\begin{equation}
\label{TwoPoint1}
\langle x(t)x(0) \rangle \equiv \langle X(t,r_{b})X(0,r_{b}) \rangle=\frac{2\alpha' e^{-\frac{k}{r_{h}^{2}}}}{r^{2}_{h}}\int_{0}^{\infty}\frac{d\omega}{\omega}
\left( \frac{2\cos(\omega t)}{e^{\beta\omega}-1}
+ e^{-i\omega  t}\right)\,, 
\end{equation}
\noindent where we have approximated the sum by an integral considering $d\omega\sim \Delta\omega$. Analogously one has 
\begin{equation}
\label{TwoPoint2}
\langle x(0)x(t) \rangle=\frac{2\alpha' e^{-\frac{k}{r_{h}^{2}}}}{r^{2}_{h}}\int_{0}^{\infty}\frac{d\omega}{\omega}
\left( \frac{2\cos(\omega t)}{e^{\beta\omega}-1}
+ e^{i\omega  t}\right) = \langle x(t)x(0) \rangle^\ast
\end{equation}
\noindent and
\begin{align}
\label{eq:CorrelationPositionTime}
\langle x(t)x(t) \rangle= \frac{2\alpha' e^{-\frac{k}{r_{h}^{2}}}}{r^{2}_{h}}\int_{0}^{\infty}\frac{d\omega}{\omega}\left( \frac{2}{e^{\beta\omega}-1}
+ 1\right) = \langle x(0)x(0) \rangle \,,
\end{align}
where these integrals are divergent, as well as the  mean square displacement. Using the normal ordering prescription, the regularized mean square displacement can be written as:
\begin{eqnarray}
    s^2_{\rm reg}(t) &\equiv& \langle : [x(t) - x(0)]^2 : \rangle \cr 
    &=&\langle :x(t)x(t):\rangle^{2} + \langle :x(0)x(0):\rangle^{2} - \langle :x(t)x(0):\rangle- \langle :x(0)x(t):\rangle \label{meanreg}
    \end{eqnarray}
Then, one gets:
\begin{equation}\label{disp2}
s^2_{\rm reg}(t) = \frac{16\alpha' e^{-\frac{k}{r_{h}^{2}}}}{r^{2}_{h}}\int_{0}^{\infty}\frac{d\omega}{\omega}\frac{\sin^{2}\left(\frac{\omega t}{2}\right)}{e^{\beta \omega}- 1}
=\frac{ 4\alpha'e^{-\frac{k}{r_{h}^{2}}} }{r^{2}_{h}}  \log \left(\frac{\sinh (\frac{t \pi}{\beta})}{\frac{t \pi}{\beta}} \right).
\end{equation}
Considering the late time approximation $t \gg \beta/\pi$, we get: 
\begin{equation}\label{latetime}
    s^2_{\rm reg}(t) \approx \frac{4\alpha'e^{-x}}{\pi T}g(x)^{2}\, t = 2Dt\,. 
\end{equation}
which is identified with the diffusive regime since  $s^{2}_{reg}\sim 2D t$. The diffusion coefficient $D$ obtained here coincides with the one given by  Eq.~\eqref{DCoefficient}, coming from the imaginary part of the admittance. 
The factor 2 in this equation is a characteristic of a one dimensional problem. On the other hand, for the short time approximation $t\ll\beta/\pi$, one finds
\begin{equation}
   s^{2}_{reg} (t) \approx  \frac{2 \alpha'e^{-x} g(x)^{2}}{3}\, t^{2} \,.
\end{equation}
which corresponds to the ballistic regime since 
$s^{2}_{reg}\sim t^2$. 

Finally, we are going to verify explicitly the  fluctuation-dissipation theorem in our holographic set up. It can be stated as \cite{Grabert:1988yt}: 
\begin{equation}\label{theorem}
  G_{\rm sym}(\omega)\equiv \frac{1}{2}\left[\langle x(\omega)x(0) \rangle+\langle x(0)x(\omega) \rangle\right]=(2n_{\rm B}+1)\rm Im(\chi(\omega)),
\end{equation}
where $G_{\rm sym}(\omega)$ is the symmetric Green's function in Fourier space and $n_{\rm B}=(e^{\beta\omega}-1)^{-1}$ is the Bose-Einstein distribution associated with thermal noise effects. 

Then, one can write the corresponding symmetric {\sl time dependent} Green's  function, using Eqs. \eqref{TwoPoint1} and \eqref{TwoPoint2}, as: 
\begin{eqnarray}
    G_{\rm sym}(t)
    &=&\frac{2\pi\alpha' e^{-\frac{k}{r_{h}^{2}}}}{r^{2}_{h}}\left(\frac{1}{2\pi}\int_{-\infty}^{\infty} d\omega \frac{2e^{-i\omega t}}{\left|\omega\right|(e^{\beta \left|\omega\right|}-1)}+\frac{1}{2\pi}\int_{-\infty}^{\infty}d\omega\frac{e^{-i\omega t}}{\left|\omega\right|}\right)\,. 
\end{eqnarray}
\noindent So, the symmetric Green's function in Fourier space is found to be: 
\begin{equation}\label{Gsym}
    G_{\rm sym}(\omega)=\left(2n_{\rm B}(\omega)+1\right)\frac{2\pi\alpha' e^{-\frac{k}{r_{h}^{2}}}}{r^{2}_{h}\omega}\,. 
\end{equation}
Furthermore, the imaginary part of the admittance $\chi(\omega)$, given by Eq. \eqref{Admittance}, is 
\begin{equation}\label{Imchi}
    {\rm Im}(\chi(\omega)) = \frac{2\pi\alpha' e^{-\frac{k}{r_{h}^{2}}}}{r^{2}_{h}\omega}
\,, 
\end{equation}
so that our deformed and backreacted model satisfies the well known fluctuation-dissipation theorem defined in Eq. \eqref{theorem}.

\section{Conclusions}\label{conc}

Here, in this work, taking into account a conformal exponential factor $\exp(k/r^2)$ and the horizon function obtained from the solutions of Einstein-dilaton equations we have constructed a deformed and backreacted Lorentz invariant holographic model. It is important to remark that the string and brane considered are in the probe approximation, so that we have disregarded their contribution to the backreaction on the metric. This means that the backreaction contribution considered here comes only from the exponential factor in the metric.

By using our  model we could investigate the fluctuation and dissipation of a string in this set up. In particular, we computed the response function (admittance), the diffusion coefficient, the relevant two-point functions and the regularized mean square displacement. From this last result we obtained the diffusive and the ballistic regimes characteristics of the Brownian motion. We also verified the fluctuation-dissipation theorem within our model from the  two-point functions and the imaginary part of the admittance. 
This analysis can be thought as an extension of the ones described in Refs. \cite{deBoer:2008gu, Tong:2012nf, Edalati:2012tc, Giataganas:2018ekx, Caldeira:2020sot}. 

The backreacted horizon function, Eq. \eqref{horfunfinal}, is displayed in Fig. \ref{hor_mu_zero} for $k\pm 1$ and $z_h=1$, where we clearly see the difference between these two choices, although they merge for low values of the holographic coordinate $z$ and also meet at $z=z_h$, satifying the condition $f(z_h)=0$. Remember that $z=1/r$, where $r$ is the radial holographic coordinate pointing outwards the black hole, so that the interval $0 < z < z_h$ represents the region outside the horizon.

The backreaction effects on the fluctuation and dissipation of the string are encoded in the function $g(x)$, defined in Eq. \eqref{HawkingTemperature}, where $x=k/r_h^2$ and $k$ is the IR scale of the model.   
This function corresponds to the deviation from the Hawking temperature due to the deformation $\exp(k/r^2)$ and the backreaction in our model with respect to the pure AdS-Schwarzschild case. In the left panel of Fig.  \ref{fig:resumo}, we show the shape of this function where one notes the asymmetry between the two branches identified with $k<0$ and $k>0$.  At $k=0$ and finite $r_h$, $g(x)|_{x=0}=1$, there is no deformation or backreaction and the Hawking temperature reduces to its usual form, $T=r_h/\pi$. 
For the branch $k<0$ the function $g(x)$ grows exponentially with $|x|\to\infty$, so the deviation from the AdS-Schwarzschild becomes larger with larger $|x|$. On the other hand, for $k>0$, $g(x)$ decreases exponentially with $x\to\infty$, vanishing for very large $x$.  

In particular, the backreaction effect on the imaginary part of the admittance is shown in the right panel of Fig. \ref{fig:resumo}. From this picture, we see that for $k<0$ 
the deviation from the pure  AdS-Schwarzschild and the deformed with no backreaction cases increases with increasing $x$. On the other side, for $k>0$ we note that the deviation from the pure AdS-Schwarzschild is limited and the two deformed solutions with or without backreaction vanish for high $x$. 

Note that the admittance $\chi(\omega)$  found here within this model  could be  compared with the ones computed from a deformed AdS space model without backreaction $\chi_{_{NBR}}(\omega)$ \cite{Caldeira:2020sot}, and the one in a geometry which includes the pure AdS-Schwarzschild case  $\chi_{_{AdS}}(\omega)$ \cite{Tong:2012nf, Edalati:2012tc, Giataganas:2018ekx}, as given by  Eq. \eqref{comparison}.

Analogously, the diffusion coefficient $D$, Eq. \eqref{DCoefficient}, obtained from  the admittance is also modified by the deformation exponential and the backreaction effects by a factor  $e^{-x}g(x)^2$. This result was checked in the calculation of the regularized mean square displacement $s^2_{reg}(t)$ from the two-point functions in the limit of late times, Eq. \eqref{latetime}. This result can be interpreted as a check of the fluctuation-dissipation theorem, Eq. \eqref{theorem}, which we verify explicitly in Eqs. \eqref{Gsym}-\eqref{Imchi}.

At this point it is appropriate to mention that Ref. \cite{Giataganas:2018ekx} extended and generalized the models discussed in Refs. \cite{Tong:2012nf, Edalati:2012tc} obtaining, from a polynomial metric, various observables related to Brownian motion as discussed in this work. Since our model is based on a conformally deformed metric, which is asymptotically AdS, both approaches can be related by a regular exponential factor $\exp{(k/r_h^2)}$. In this sense, some of our results, as the admittance Eq. \eqref{Admittance}, could have been inferred from the work \cite{Giataganas:2018ekx}.  

As a last comment, we would like to highlight the main result of this work. The effect of the backreaction considered here enhances the results on the physical quantities related to the fluctuation and dissipation of the string presented in  Ref. \cite{Caldeira:2020sot}.

\begin{acknowledgments}

The authors would like to thank Diego M. Rodrigues and Alfonso Ballon-Bayona for discussions.  N.G.C. is supported by  Conselho Nacional de Desenvolvimento Científico e Tecnológico (CNPq) and Coordenação de Aperfeiçoamento de Pessoal de Nível Superior (CAPES). H.B.-F. and C.A.D.Z. are partially supported by Conselho Nacional de Desenvolvimento Cient\'{\i}fico e Tecnol\'{o}gico (CNPq) under grants No. 311079/2019-9 and No.309982/2018-9, respectively.

\end{acknowledgments}

\end{document}